# The Formation of Self-Assembled Nanowire Arrays on Ge(001): a DFT Study of Pt Induced Nanowire Arrays

Danny E. P. Vanpoucke[1] and Geert Brocks[1]
[1]Computational Materials Science, University of Twente, P.O. Box 217,
7500 AE Enschede, The Netherlands

**ABSTRACT**

Nanowire (NW) arrays form spontaneously after high temperature annealing of a sub monolayer deposition of Pt on a Ge(001) surface. These NWs are a single atom wide, with a length limited only by the underlying beta-terrace to which they are uniquely connected. Using *ab-initio* density functional theory (DFT) calculations we study possible geometries of the NWs and substrate. Direct comparison to experiment is made via calculated scanning tunneling microscope (STM) images. Based on these images, geometries for the beta-terrace and the NWs are identified, and a formation path for the nanowires as function of increasing local Pt density is presented. We show the beta-terrace to be a dimer row surface reconstruction with a checkerboard pattern of Ge-Ge and Pt-Ge dimers. Most remarkably, comparison of calculated to experimental STM images shows the NWs to consist of *germanium* atoms embedded in the Pt-lined troughs of the underlying surface, contrary to what was assumed previously in experiments.

**INTRODUCTION**

Microelectronics industry is driven by the ever further miniaturization of electronic components and devices. Monatomic nanowires present the physical limits of miniaturization for wires, and much research interest is directed toward the construction of wires closer to this physical limit and understanding the associated physical phenomena.
In 2003, Gürlü *et al*. [1] observed self-assembled nanowire arrays after deposition of 0.25 monolayer (ML) of Pt on a clean Ge(001) surface and subsequent high temperature annealing. The NWs are equally spaced at 1.6 nm, virtually defect and kink free, and only a single atom wide while being hundreds of nanometers long. The NWs, dubbed Pt-nanowires, turn out not to be conductive. Instead the regions between the NWs show conductive behavior [2].
Using *ab-initio* DFT calculations we study possible geometries for the beta-terrace and the NW arrays. Direct comparison to experiment was made using calculated STM images, allowing for the identification of the geometry of the experimentally observed structures.

**THEORETICAL METHOD**

DFT calculations are performed using the projector augmented wave (PAW) method and local density approximation (LDA) functional as implemented in the Vienna Ab-initio Simulation Package (VASP) [3-6]. A plane wave basis set with kinetic energy cutoff of 345 eV is applied and the surface Brillouin zone is sampled using an 8x4x1 Monkhorst-Pack special k-point mesh. Symmetric slabs of 12 layers, separated by 1.55 nm of vacuum, are used with reconstructions on both surfaces. The geometries are optimized using the conjugate gradient method. The center two layers of the slab are kept fixed to represent bulk conditions.

STM images are calculated using the Tersoff-Hamann method [7], which states that the tunneling current in an STM experiment is proportional to the local density of states. It is used in its simplest form, with the STM-tip approximated as point-source. Experimental surfaces of constant current are simulated through surfaces of constant density, representing the constant current mode in an STM experiment [8].

**RESULTS AND DISCUSSION**

The observed NWs on Ge(001) are inherently connected to the beta-terrace. This makes a study of the beta-terrace a necessary first step toward understanding the formation of the self-assembled NWs on Ge(001). The experimental STM image of the beta-terrace in Figure 1a clearly show the presence of dimer rows with a c(4x2) symmetry, consisting of two different types of dimers. In experiment about 0.25 ML of Pt was deposited on a clean Ge(001) surface, and after high temperature annealing three types of terraces were observed: The alpha-terrace, showing a Ge(001) surface reconstruction riddled with dimer vacancies, the beta-terrace, showing dimer rows with two types of dimers, and the NW arrays, which are always observed on a beta-terrace. The presence of the three types of terraces suggests that the Pt is not distributed homogeneously over the surface.

**<u>Beta-terrace</u>**

We start from the assumption that the beta-terrace contains 0.25 ML of Pt, or 2 Pt atoms per 4x2 surface unit cell. Two Ge atoms in the first layer were substituted by Pt atoms, one at position 0 and the other at one of the remaining 7 positions, giving rise to seven different geometries: beta1 to beta7 (see Figure 2a). Taking into account the buckling of the surface dimers and the symmetry, the number of different geometries increases to ten.
For all these structures the formation energy, per surface unit cell, $E_f$ is calculated using

$$E_f = E_{rec} - E_{Ge(001)} - N_{Pt}(E_{Ge} - E_{Pt}), \tag{1}$$

with $E_{rec}$ the total energy of the studied structure, $E_{Ge(001)}$ the total energy of the asymmetric (2x1) reconstructed Ge(001) surface, $N_{Pt}$ the number of Pt atoms in the system and $E_{Ge}$ and $E_{Pt}$ the bulk energy of a Ge and Pt atom respectively. Negative values of the formation energy indicate an increase in stability of the system with regard to a reconstructed Ge(001) surface and bulk Pt.

The structure with the embedded Pt dimer (beta1) turns out to be very unstable with a formation energy that is 0.4-0.7eV per surface unit cell higher than of the other structures. For all other structures the formation energies per Pt-Ge surface dimer lie within a range of 150 meV, making them accessible at the annealing temperature of >1000 K.

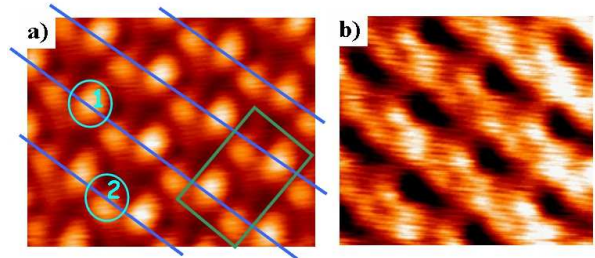

**Figure 1.** Experimental filled (a) and empty (b) state STM images of the beta-terrace. Used Bias voltage : -0.3V (a) and +0.3V (b) [9]. The lines indicate the direction of the dimer rows in the filled state image. The rectangle shows the c(4x2) surface unit cell, and the two types of dimers are indicated with ellipses. The most intriguing feature of the empty state image are the triangular holes between the dimer rows.

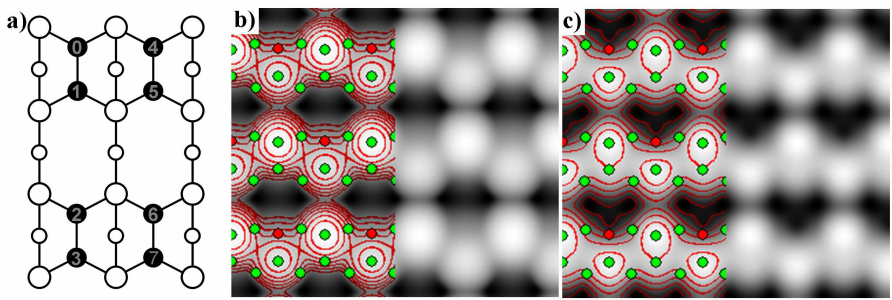

**Figure 2.** a) Schematic representation of a reconstructed Ge(001) surface. Calculated filled (b) and empty (c) state STM images of the beta6-geometry at simulated bias of -0.70V (b) and +0.70V(c). Green(red) discs indicate the positions of the Ge(Pt) atoms in the two top-layers of the surface. Contours are added to guide the eye.

STM images are generated for all the structures, and it turns out that the filled state images can composed of 3 different dimer images, each connected to a specific dimer geometry. A bright big round image for the Ge-Ge dimer. A bright rectangular image for the Pt-Ge dimer with the Pt atom at the down side of the dimer. Finally, a smaller dim round dimer for the Pt-Ge dimer with Pt at the up side of the dimer. Comparison to the experimental image then allows the identification of the geometry of the beta-terrace, on the basis of simple elimination.

1) The beta-terrace has 2 types of dimers: elimination of beta3, beta5 and beta7
2) The beta-terrace has a c(4x2) symmetry: elimination of beta1, beta2 and beta4

The only remaining structure is the beta6 geometry, which has the second most favorable formation energy $E_f$ = -0.05 eV (beta4 being the most favorable with $E_f$ =-0.12 eV). The calculated filled state STM images show a good agreement with the experimental image. This also allows us to identify the bright dimers in experiment as Ge-Ge dimers, while the darker dimers are Pt-Ge dimers, with the Pt atom at the dark side of the dimer (see Figures 1a and 2b). The empty state images also show very good agreement. In addition, the experimentally observed triangular dark areas are only present in the calculated STM images of the beta6 geometry (see Figures 1b and 2c)

**Nanowires**

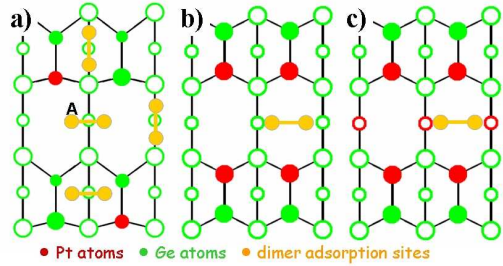

**Figure 3.** Schematic representation of the beta6-geometry (a) and two models for the NW substrate (b and c) with additional Pt in the substrate. Possible adsorption sites for Pt/Ge dimers are indicated. The same adsorption sites as for the beta6-geometry are used for the Ge(001) surface.

Starting from the structure found for the beta-terrace, two simple scenarios for NW formation could be imagined. In the first scenario the Pt atoms of the beta-terrace are ejected onto the surface during annealing and form dimers in the troughs, thus creating the NW arrays. The vacancies in the top layer are then filled with Ge atoms from the bulk, in the course of which a normal Ge(001) surface is recovered. In this scenario a homogeneous Pt distribution of 0.25 ML is maintained. In the second scenario, the Pt atoms in the top layer of the beta-terrace remain at their positions, but extra Pt atoms located deeper into the bulk are ejected onto the surface during annealing, and form dimers in the troughs to create NW arrays. For both scenarios several adsorption structures, shown in figure 3a, are examined, and the binding energy $E_b$ is calculated using the equation:

$$E_b = (E_{rec} - E_{substrate} - 2N_{ad}E_X)/N_{ad}, \qquad (2)$$

with $E_{substrate}$ the total energy of the substrate structure, $N_{ad}$ the number of ad-dimers per surface unit cell and $E_X$ the bulk energy for the atomic species of the adsorbed dimers. Again, a negative $E_b$ value indicates an increase of the stability, whereas a positive value indicates a decrease.

The nanowire binding energies show the adsorption of Pt dimers on the Ge(001) and beta6-surface to be very unstable. $E_b = 2.3$ eV per Pt dimer, for a Pt dimer adsorbed on a dimer row of a Ge(001) surface. Pt dimers positioned in the troughs are about 2 eV more stable but still unstable in the absolute sense. Pt dimers adsorbed on the beta6-surface often cause large deformations of the surface, with Pt atoms moving into subsurface positions. Calculated STM images of these structures reveal only one structure vaguely resembling a NW. This is the Pt dimer adsorbed in the trough, parallel to the dimer row of a Ge(001) surface (see Figure 3a, site A). The Pt atoms bind to Ge atoms of the surface dimers, which show up as bright dimers in the calculated STM images. Despite the resemblance to the experimental NWs some crucial differences are present. Besides having an unfavorable binding energy of ~0.4 eV per Pt dimer, the NW image is also located on top of the dimer row, in contradiction to experiment. Furthermore, specific image details, like the symmetric bulges are missing. However, this structure gives an important insight, namely, that the experimentally observed NW could consist of Ge atoms, since these are the atoms showing bright images in calculates STM pictures.

Based on these results a second surface model is build (see Figure 3b), containing 0.5 ML of Pt in the top layer. Pt and Ge dimers are placed in the Pt-lined trough and the formation and binding energies are calculated. The binding energy for the adsorbed Ge(Pt) dimer is -0.76(-1.25) eV per dimer. Although the Pt dimer is most stable, this structure presents a problem. The calculated STM images of this structure show no NWs. At the position of the NW only a widened trough is visible (see Figure 4a and e) in the image. Also the dimer row images have changed. Only one feature for every two dimers is visible in the image. However, this image of a Pt NW adsorbed on this substrate model, shows very good agreement with the experimentally observed widened troughs, which are considered a precursor to the formation of the NWs [10]. In contrast, the calculated STM images of an adsorbed Ge NW show a very clear NW image, displaying most of the features observed in experimental NWs.

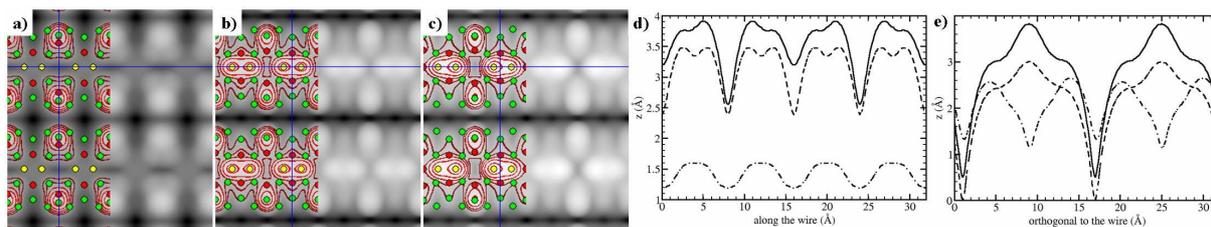

**Figure 4.** Calculated filled state STM images at a simulated bias of -1.5V, of (a) Pt NW adsorbed on the substrate shown in Figure 3b, (b) Ge NW adsorbed on the substrate shown in figure 3c, and (c) the Ge NW with extra Pt atom in the trough presenting a 4x1 periodicity. The green/red/yellow discs show the positions of the Ge/Pt/NW-atoms in these structures. Cross-sections of the STM images along (d) and orthogonal (e) to the wire direction are given for the 3 systems: dash-dotted line for the Pt NW (a), dashed line for the Ge NW (b) and solid line for the Ge NW (c).

Geometrically, the Pt NW sinks into the trough while the Ge NW stays above it. This leads to a third surface model (see Figure 3c), where the Ge atoms at the bottom of the Pt lined trough are replaced by Pt atoms. On this surface the adsorption of a Pt dimer is slightly unstable, while the binding energy of the Ge dimer is -1.04 eV. Furthermore, going from the structure with the Pt dimer discussed in the previous paragraph to this Ge dimer structure can be done by the exchange of the Pt dimer with the Ge atoms at the bottom of the trough just underneath this Pt dimer. The gain in formation energy by this exchange is 0.33 eV per dimer, indicating that the spontaneous transformation from the widened trough to a Ge NW is energetically favorable. The calculated STM images of this Ge NW show perfect agreement with the experimentally observed NWs (see Figure 4b). The cross-section of the calculated STM image along the NW shows clearly the presence of double peaked dimer images in the filled states (see Figure 4d).

Experimentally one can observe a 4x1 periodicity along a NW, but only for those NWs that are situated on the inside of a patch of NWs. Our calculations do not show this because of the size of our unit cell, which is only half the length needed for this periodicity. Calculations for a doubled cell however show no spontaneous appearance of a 4x1 periodicity along the NW. The NW dimers remain horizontal. Even geometries starting with tilted dimers relax back to horizontal NW dimers. This would lead to the conclusion that the 4x1 periodicity is not caused by a Peierls instability. When extra Pt and/or Ge atoms are added to the trough, we found that by adding 1 Pt atom per 2 Ge NW dimers a 4x1 periodicity can be attained (see Figure 4c). This extra Pt atom resides at the bottom of the trough, and binds to two Ge-NW-dimers. This pins both dimers on their location and stabilizes the structure. It also pulls the dimers slightly inward tilting them over 3.6°, resulting in the observed 4x1 periodicity along the NW.

# CONCLUSIONS

We have presented the structure of the beta-terrace. It is a Ge (001) c(4x2) surface reconstruction containing 0.25 ML of Pt in the top layer. The checkerboard pattern is formed by bright Ge-Ge dimers and dim Pt-Ge dimers. The substrate for the NWs is not a simple beta-terrace, but a surface structure containing at least 0.75 ML of Pt in the top layers. Pt NWs are invisible in calculated STM images and appear as widened troughs. When these Pt atoms exchange position with the Ge atoms underneath, the latter form dimers giving rise to the appearance of NW arrays in STM images. We conclude that the experimentally observed NWs are composed of "*germanium dimers*" embedded in Pt lined troughs. Inside the NW arrays where the local Pt density is slightly higher, extra Pt atoms in the Pt lined troughs induce a 4x1 periodicity along the Ge NW, making this periodicity doubling a purely structural effect.

With regard to the three types of experimentally observed terraces, this leads to the identification of two surface phase-transitions. The first at a local Pt density of 0.25 ML, when an alpha-terrace becomes a beta-terrace, and a second at just slightly over 0.75 ML, where the beta-terrace transforms into a Ge NW array.

# ACKNOWLEDGMENTS

The authors would like to thank Prof. H. J. W. Zandvliet and Dr. A. van Houselt for making their experimental STM results available. This work is part of the research program of the "Stichting voor Fundamenteel Onderzoek der Materie" (FOM) and the use of supercomputer facilities was sponsored by the "Stichting Nationale Computer Faciliteiten" (NCF), both financially supported by the "Nederlandse Organisatie voor Wetenschappelijk Onderzoek" (NWO).